\documentstyle[prabib,aps,psfig,12pt]{revtex}

\begin{document}
\title{Nondifferentiable Dynamic: Two Examples}
\author{V. Dzhunushaliev
\thanks{E-Mail Addresses : dzhun@rz.uni-potsdam.de and 
dzhun@freenet.bishkek.su; \qquad permanent address: 
Dept. Theor. Phys., Kyrgyz State National University, 
Bishkek 720024, Kyrgyzstan}}
\address{Institut f\"ur Mathematik, Universit\"at Potsdam 
PF 601553, D-14415 Potsdam, Germany} 

\maketitle

\vspace{1cm}
\begin{abstract}
Some nondifferentiable quantities (for example, the 
metric signature) can be the independent physical degrees 
of freedom. It is supposed that in quantum gravity these 
degrees of freedom can fluctuate. Two examples of such 
quantum fluctuation are considered: a quantum 
interchange of the sign of two components of the 5D 
metric and a quantum fluctuation between Euclidean and 
Lorentzian metrics. The first case leads to a spin-like 
structure on the throat of composite wormhole and to a possible 
inner structure of the string. The second case leads to 
a quantum birth of the non-singular Euclidean Universe with frozen 
$5^{th}$ dimension. The probability for such quantum 
fluctuations is connected with an algorithmical complexity 
of the Einstein equations. 
 
\end{abstract}

\section{Introduction}

It is possible that some discrete mathematical objects 
which can not be the continuous functions nevertheless are 
the physical degrees of freedom in the Nature. It can be 
a signature of metric, dimensionality, topology of space and 
so on. A time evolution of such kind of the variables is a 
big problem for the classical and quantum gravity. 
\par 
The change of the metric signature in the classical gravity ordinary is 
connected with the presence of a surface $\delta$-like 
matter (see for example, \cite{ellis95}, \cite{ellis97}). 
Certainly, 
we have the question: is this matter exotic or ordinary, 
i.e. can such conditions realized in the Nature ? 
In quantum gravity the change of metric signature is the 
result of integrating in the path integral over the 
Euclidean and Lorentzian metrics. The difficulties connected 
with this problem is easy to see in the vier-bein formalism 
\begin{equation}
ds^2_{(5)} = \eta _{ab}e^a e^b
\label{in-1}
\end{equation}
here $\eta_{ab} = (-,+,+,+)$, $e^a = h^a_\mu dx^\mu$, 
$a = 0,1,2,3$ is the vier-bein index, $\mu$ is the spacetime 
index. In the classical regime only the tetrad 
components $h^a_\mu$ is the dynamical variables and varying 
with respect to $h^a_\mu$ leads to the Einstein equations. 
But we can not vary with respect to $\eta_{ab}$ 
and therefore we have not the corresponding equations. 
This allow us to say that the difficulties connected 
with the signature change are connected with that $\eta_{ab}$ 
are the nondynamical variables. In quantum gravity $\eta_{ab}$ 
become the dynamical quantities. We see that $\eta_{ab}$ are the 
discrete variables and in fact an integration over $\eta_{ab}$ 
should be a summation. 
\par 
In this paper we would like to show that in quantum gravity 
can exist some discrete (nondifferentiable) physical degrees 
of freedom\footnote{here we consider the case with the components 
of $\eta _{ab}$}
which also can be the dynamical variables. 
In particular the signature change can happen not on the boundary 
between different regions with the Euclidean and Lorentzian 
regions but it can take place a fluctuation (``quantum trembling'') 
between $\eta _{ab} = +1$ and $\eta _{ab} = -1$. 

\section{Fluctuation without the change of the sign 
of metric determinant}
\label{without}

Let we consider the 5D metric
\begin{equation}
ds^{2} = - \sigma \Delta (r)dt^{2} + 
dr^{2} + a(r)d\Omega ^2 + 
\sigma \frac{r_1^2}{\Delta (r)}(d\chi - \omega (r)dt)^2 
\label{tr-1}
\end{equation}
here $\chi $ is the 5$^{th}$ extra coordinate; 
$r,\theta ,\varphi $ are the $3D$  polar coordinates; 
$t$ is the time; $d\Omega ^2 = d\theta ^{2} + \sin ^{2}\theta  d\varphi ^2$ 
is the metric on the $S^2$ sphere; $\sigma = \pm 1$ describes 
the interchange of metric signature: 
$(-,+,+,+,+) \leftrightarrow (+,+,+,+,-)$. 
The functions $\Delta (r), a(r)$ are 
the even functions this means that the 3D part of the 
(\ref{tr-1}) metric is the wormhole-like 3D space. The 
5D vacuum Einstein equations give us 
\begin{eqnarray}
\frac{\Delta ''}{\Delta} - \frac{{\Delta '}^2}{\Delta ^2} +  
\frac{a' \Delta '}{a\Delta} - \frac{r_0^2}{\Delta ^2}{\omega '}^2 & = & 0 ,
\label{tr-2}\\
\omega '' - 2 \omega ' \frac{\Delta '}{\Delta} + 
\omega ' \frac{a'}{a} & = & 0 ,
\label{tr-3}\\
\frac{{\Delta '}^2}{\Delta ^2} + \frac{4}{a} - 
\frac{{a'}^2}{a^2} - \frac{r_0^2}{\Delta ^2}{\omega '}^2 & = & 0 ,
\label{tr-4}\\
a'' - 2& = & 0 .
\label{2-4}
\end{eqnarray}
with the following solution \cite{dzh2} 
\begin{eqnarray}
a & = & r^{2}_{0} + r^{2},
\label{tr-5}\\
\Delta & = & \frac{2r_0}{q}\frac{r^2 + r_0^2}
{r^2 - r_0^2} ,
\label{tr-6}\\
\omega & = &  \frac{4r_0^2}{r_1q}\frac{r}
{r^2 - r_0^2} .
\label{tr-7}
\end{eqnarray}
here $r_0 > 0$ and $q$ are some constants. 
We see that the (\ref{tr-2})-(\ref{tr-4}) equations 
do not depend on the $\sigma$. 
It is the most important thing for understanding 
as occurs a quantum fluctuation of the metric signature. 
We have one solution for two metrics with the different 
signature and the classical dynamical 
equations (\ref{tr-2})-(\ref{tr-4}) can not distinguish 
them. But the quantum paradigm says us: 
\textit{that which is not forbidden is permitted}. Following this 
rule we can say that in this situation should exist the 
fluctuations (``quantum trembling'') between two signatures.
\par 
Let $\eta_1$ is a quantum state with the $(-,+,+,+,+)$ signature 
and $\eta_2$ with the $(+,+,+,+,-)$ signature. Then we can 
assume that 
\begin{equation}
\eta_1 = \frac{1}{\sqrt 2}\left ({1 \atop 0} \right ) , \quad 
\eta_2 = \frac{1}{\sqrt 2}\left ({0 \atop 1} \right )
\label{tr-8}
\end{equation}
as the probabilities for 
both signatures $(\pm,+,+,+,\mp)$ should be equal. 
The eigenstates $\eta _{1,2}$ describe the states with 
$(\mp ,+,+,+,\pm)$ accordingly. 
\par 
We see that the eigenstates (\ref{tr-8}) are the same 
as the eigenstates of z-component of the spin 
\begin{eqnarray}
\left (\frac{\hbar}{2} \sigma_3\right )\zeta & = & 
\pm \frac{\hbar}{2}\zeta
\label{tr-9}\\
\zeta_1 = \frac{1}{\sqrt 2}\left ({1 \atop 0} \right ) , & \quad &
\zeta_2 = \frac{1}{\sqrt 2}\left ({0 \atop 1} \right )
\label{tr-10}
\end{eqnarray}
here $\frac{\hbar}{2}\sigma_3$ is the operator of $z$-component 
of the spin, $\zeta_{1,2}$ are its eigenstates and 
$\pm \frac{\hbar}{2}$ are its eigenvalues. 
\par
\begin{figure}
\centerline{
\framebox{
\psfig{figure=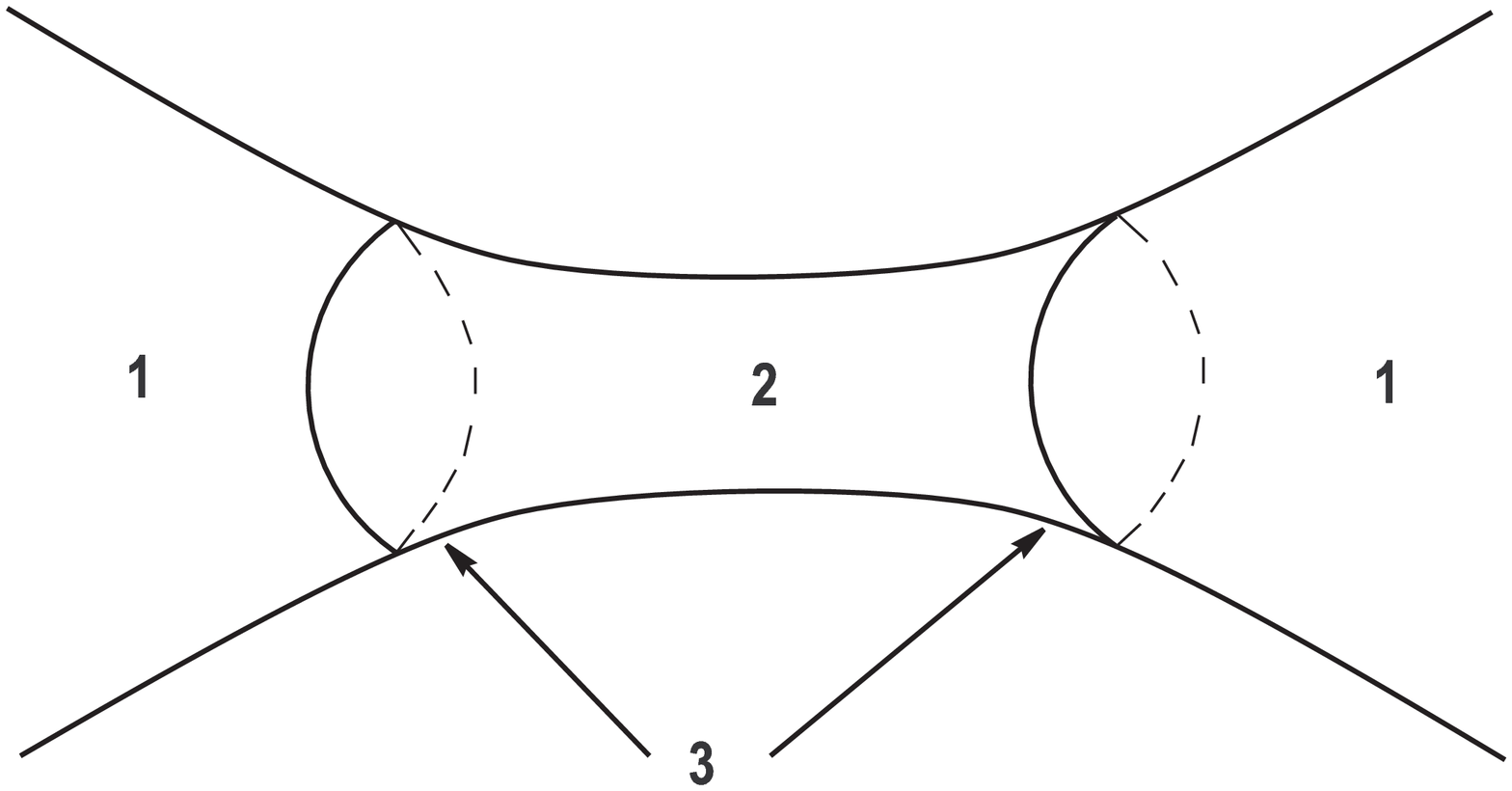,height=5cm,width=5cm,angle=-90}}}
\vspace{5mm}
\caption{The composite wormhole: whole space is 5 
dimensional but in {\bf 1} regions (black holes) 
the $G_{55}$ metric component 
is non-dynamical (in this case 5D gravity is completely 
equivalent to 4D gravity + electromagnetism), {\bf 2} region 
(5D wormhole-like solution (\ref{tr-5})-(\ref{tr-7})) 
is the 5D throat where $G_{55}$ is the dynamical variable, 
{\bf 3}-hypersurfaces are the event horizons.} 
\label{fig1}
\end{figure}

 
This allow us to presume that such ``quantum trembling'' 
between two signatures 
exists and have the physical interpretation in the 
following sense. In \cite{dzh7} it was proposed a model of 
composite wormhole consisting from a 5D throat\footnote{which 
is the solution (\ref{tr-5})-(\ref{tr-7})} 
and two Reissner-Nordstr\"om black holes attached to the 
5D throat on the event horizon, see Fig.\ref{fig1}. 
\par
In Ref. \cite{wheel2} Wheeler 
assumed that the geometrical origin of spin 
probably is connected with a quantum fluctuation 
orientability $\leftrightarrow$ nonorientability\footnote{i.e. 
we have a ``two-valuedness''}: 
\textit{
``If however both spaces\footnote{orientable and non-orientable} 
are permissible, then a space with 
one wormhole has one classical two-valuedness associated with it, 
and one with $n$ wormholes has $n$-fold duplicity. If it 
should turn out that there are $2^n$ inequivalent ways to 
get to a geometry with $n$ wormholes, and if it should make 
sense to assign $2^n$ distinct probability amplitudes 
to the same macroscopic field configuration, then one would 
be in possession of a non-classical two valuedness with as 
many spin-like degrees of freedom as there are wormholes. 
$\ldots$ A number of options shows up which is qualitatively 
of the same order as the number of degrees of freedom of 
a spinor field, when one goes to the virtual foam-like 
space of quantum geometrodynamics. It is difficult to say 
anything more specific about the reasonableness or 
unreasonableness of this conceivable "correlation of spin 
with parity" until more is known about the formalism of quantum 
geometrodynamic''}. 
Here we offer the above-mentioned ``unclassical two-valuedness'' 
connected with ``quantum trembling'' 
between two classical solutions with the different metric 
signature as a model of an inner geometrical structure of spin 
in spirit of the Wheeler idea ``spin = two-valuedness''. 
\par
\begin{figure}[htb]
\centerline{
\framebox{
\psfig{figure=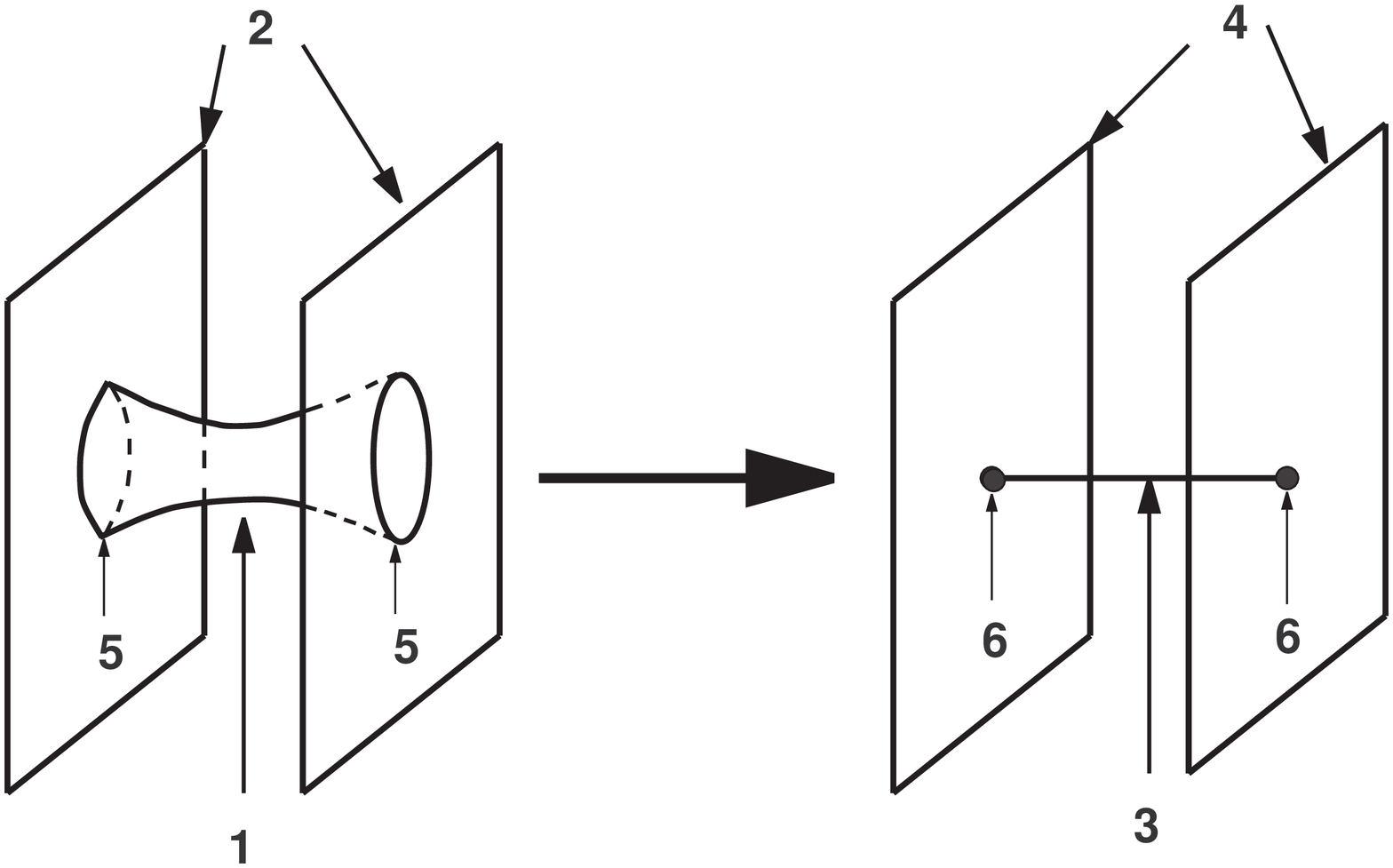,height=5cm,width=5cm}}}
\vspace{5mm}
\par
\caption{The factorization of the 5D throat-{\bf (1)} 
of composite WH 
by $S^2$ spheres leads to a string-like object-{\bf (3)} 
attached to two branes-{\bf (4)}, {\bf (2)} are two black holes. 
The event horizons-{\bf (5)} become the points-{\bf (6)} of attaching 
the string to D-branes. The throat of composite WH with the flux 
of electric field after factorization is the string 
with the flux of electric field and the electric charges 
at the ends of string. The quantum fluctuation 
$(-,+,+,+,+) \leftrightarrow (+,+,+,+,-)$ 
leads to the appearance of a spin-like (super) structure 
on the string.}
\label{fig2}
\end{figure}

 
It is very interesting that our composite WH in some 
approximation is close to a string attached to two 
D-branes. Let we factorize 
the 5$^{th}$ throat by the manner represented on the Fig.\ref{fig2}, 
i.e. all $S^2$ spheres in the left side of the picture 
are contracted to the points on the right side, two event horizons 
are contracted to two points of attachment the string to the branes. 
\par
\par
As we saw above ``quantum trembling'' between signatures 
leads to the appearance of the spin-like structure, i.e. to the 
fermion degrees of freedom. But it is well known that 
these degrees of freedom are connected with the grassmanian 
coordinates. It is possible that this means that the 
quantum fluctuations between two metric signatures leads 
to the appearance of a superspace with the ordinary and 
grassmanian coordinates. 
In this case the throat of composite WH + fluctuation 
between two signatures after above-mentioned factorization 
is equivalent to a superstring attached to two branes, 
see Fig.\ref{fig2}. The whole composite WH is equivalent 
to the D-brane (superstring + two branes). It allow us to assume 
that the superstring possible has an inner structure and in 
this sense it is an approximation for the composite WH. 
In this connection it can mention the citation from 
Ref.\cite{hooft99}: 
\textit{
``Discrete degrees 
of freedom often manifest themselves as fermion in the quantum 
formalism. It is also conceivable that the continuum 
theories at the basis of our considerations will have up include 
string- and D-brane degrees of freedom $\ldots$ ''.} 

\section{Fluctuation with the change of the sign
of metric determinant}
\label{with}

In the previous section we consider the case of the quantum 
fluctuation between two signatures 
$(-,+,+,+,+) \leftrightarrow (+,+,+,+,-)$ 
without the change of the sign of metric determinant. Here 
we assume that can be a situation when a fluctuation 
between the Euclidean and Lorentzian metrics is possible. 
Evidently this can be only a cosmological solution in contrast 
with the previous spherically symmetric case. 
This idea differs from the initial Hawking idea about 
changing of the metric signature on the boundary between 
Euclidean and Lorentzian regions in such a manner that 
in the Universe is a region where takes place a quantum 
fluctuation (``trembling'') between different metric 
signature. It can be in the very Early Universe on the level of 
Planck scale.
\par 
For example, we examine a vacuum 5D Universe with the metric 
\begin{eqnarray}
ds^2_{(5)} = \sigma dt^2 + b(t)\left (d\xi + 
\cos \theta d\varphi \right )^2 + a(r)d\Omega ^2_2 + &&  
\nonumber \\
r_0^2 e^{2\psi (t)}\left [d\chi - \omega (t) \left (d\xi + 
\cos \theta d\varphi \right ) \right ]^2 &&
\label{cs-1}
\end{eqnarray}
here $\sigma = \pm 1$ for the Euclidean and 
Lorentzian signatures respectively. 3D space metric 
$dl^2 = b(t)\left (d\xi + 
\cos \theta d\varphi \right )^2 + a(r)d\Omega ^2_2$ 
describes the Hopf bundle with the $S^1$ fibre over the 
$S^2$ base. In the 5-bein formalism we have 
\begin{equation}
ds^2_{(5)} = \eta _{\bar A\bar B} e^{\bar A}e^{\bar B}
\label{cs-2}
\end{equation}
here $\bar A, \bar B$ are the 5-bein indexes and 
\begin{eqnarray}
\eta _{\bar A\bar B} & = & \left (\pm 1,+1,+1,+1,+1 \right ) ,
\label{cs-3}\\
e^{\bar 0} & = & dt ,
\label{cs-4}\\ 
e^{\bar 1} & = & \sqrt b \left (d\xi + 
\cos\theta d\varphi\right ) ,
\label{cs-5}\\
e^{\bar 2} & = & \sqrt a d\theta ,
\label{cs-6}\\
e^{\bar 3} & = & \sqrt a \sin\theta d\varphi ,
\label{cs-7}\\
e^{\bar 5} & = & r_0 e^{\psi}
\left [d\chi - \omega (t) \left (d\xi + 
\cos \theta d\varphi \right ) \right ]
\label{cs-8}
\end{eqnarray}
According to the following theorem \cite{Sal1}, \cite{Per1}:
\par
\textit{
Let $G$ be a structural group
of the principal  bundle.  Then  there  is a one-to-one
correspondence between the $G$-invariant metrics
\begin{equation}
ds^2 = \varphi (x^\alpha) \Sigma ^a \Sigma _a +
g_{\mu\nu}(x^\alpha) dx ^\mu dx^\nu
\label{cs-9}
\end{equation}
on the  total  space ${\cal X}$
and the triples $(g_{\mu \nu }, A^{a}_{\mu }, \varphi )$.
Here $g_{\mu \nu }$ is the 4D Einstein's pseudo  -
Riemannian metric on the base; $A^{a}_{\mu }$ are the gauge fields
of the group $G$ ( the nondiagonal components of
the multidimensional metric); $\varphi \gamma _{ab}$  is the
symmetric metric on the fibre 
($\Sigma ^a = \sigma ^a + A^a_\mu (x^\alpha)dx^\mu$, 
$\Sigma _a = \gamma _{ab}\Sigma ^b$, 
$\gamma _{ab} = \delta_{ab}$; 
$a=5, \cdots , \dim G$ is the index on the fibre and 
$\mu = 0,1,2,3$ is the index on the base).}
\par
we have the electromagnetic potential
\begin{equation}
A = \omega (t) \left (d\xi + 
\cos \theta d\varphi \right ) = 
\frac{\omega}{\sqrt b} e^{\bar 1}
\label{cs-10}
\end{equation}
For this potential the Maxwell tensor is
\begin{equation}
F = dA = \frac{\dot \omega}{\sqrt b} e^{\bar 0} \wedge e^{\bar 1} - 
\frac{\omega}{a}e^{\bar 2} \wedge e^{\bar 3}
\label{cs-11}
\end{equation}
Therefore we have the electrical field
\begin{equation}
E_{\bar 1} = F_{\bar 0 \bar 1} =  \frac{\dot \omega}{\sqrt b}
\label{cs-12}
\end{equation}
and the magnetic field
\begin{equation}
H_{\bar 1} = \frac{1}{2}\epsilon_{1\bar j\bar k}F^{\bar j \bar k} = 
- \frac{\omega}{a}
\label{cs-13}
\end{equation}
Let we write down the vacuum 5D Einstein equations
\begin{eqnarray}
G_{\bar 0\bar 0} \propto 
2\frac{\dot b \dot \psi}{b} + 4 \frac{\dot a \dot \psi}{a} + 
2\frac{\dot a \dot b}{ab} + \frac{\dot a^2}{a^2} + 
\sigma \left (\frac{b}{a^2} - \frac{4}{a}\right ) + 
r_0^2e^{2\psi}\left (\sigma H_{\bar 1}^2 - E_{\bar 1}^2 
\right ) & = & 0 ,
\label{cs-14}\\
G_{\bar 1\bar 1} \propto 4\ddot\psi + 4{\dot\psi}^2 + 
4\frac{\ddot a}{a} + 4\frac{\dot a\dot\psi}{a} + 
\sigma\left (3\frac{b}{a^2} - \frac{4}{a} \right ) - 
\frac{{\dot a}^2}{a^2} + 
r_0^2e^{2\psi}\left (\sigma H_{\bar 1}^2 - E_{\bar 1}^2 
\right ) & = & 0 ,
\label{cs-15}\\
G_{\bar 2\bar 2} = G_{\bar 3\bar 3} \propto 
\nonumber\\
4\ddot\psi + 4{\dot\psi}^2 + 
2\frac{\ddot b}{b} + 2\frac{\dot b\dot\psi}{b} - 
\frac{\dot b^2}{b^2} + 2\frac{\ddot a}{a} + 
2\frac{\dot a\dot\psi}{a} + 
\frac{\dot a\dot b}{ab} - \frac{\dot a^2}{a^2} - 
\sigma\frac{b}{a^2} - 
r_0^2e^{2\psi}\left (\sigma H_{\bar 1}^2 - E_{\bar 1}^2 
\right ) & = & 0 ,
\label{cs-16}\\
R_{\bar 5\bar 5} \propto \ddot \psi + {\dot\psi}^2 + 
\frac{\dot a\dot\psi}{a} + \frac{\dot b\dot\psi}{2b} + 
\frac{r_0^2}{2}e^{2\psi}\left (
\sigma H_{\bar 1}^2 + E_{\bar 1}^2
\right ) &= & 0 ,
\label{cs-17}\\
R_{\bar 2\bar 5} \propto \ddot \omega + \dot\omega
\left (\frac{\dot a}{a} - \frac{\dot b}{2b} + 3\dot\psi
 \right )
-\sigma \frac{b}{a^2}\omega & = & 0
\label{cs-18}
\end{eqnarray}
where 
$G_{\bar A\bar B} = R_{\bar A\bar B} - 
\frac{1}{2}\eta_{\bar A\bar B}R$ is the Einstein tensor. 
\par 
Now we can formulate our basic assumption: 
\textit{by some conditions, in one region\footnote{for 
example, in the very Early 
Universe}, can exist a quantum fluctuation between the 
Euclidean and Lorentzian metric signatures}. This means that 
in the classical equations (\ref{cs-14})-(\ref{cs-18}) 
arises a quantum fluctuating quantity $\sigma$ defining 
the metric signature. Another words, we have two copies of the 
classical equations: one with $\sigma = +1$ and another 
with $\sigma = -1$. The equation (\ref{cs-14}) is invariant 
relative to $\sigma = \pm 1$ exchange. Let we consider the 
remaining equations with $\sigma$. The basic question arising 
in this situation is: how is a probability for each equation 
with $\sigma = +1$ and $\sigma = -1$ in the 
(\ref{cs-15})-(\ref{cs-18}) equations system ? 
\par
As in Ref. \cite{dzh4} we will define this probability 
starting from the algorithmical complexity (AC) of 
each equation. What is the AC and how is its physical 
interpretation? In 60$^{th}$ Kolmogorov have defined 
the notion of probability from the algorithmical point 
of view \cite{kol1}. His basic idea is very simple: 
a probability of an appearance of some object depends 
from its AC and the AC is defined 
as the \textit{minimal} length of an algorithm describing 
given object on some universal computer (Turing machine, 
for example). Simply speaking: \textit{the simpler the more 
probable}. 
\par
The key word for such definition of the probability 
is the word ``minimal''. In this case the length of the 
algorithm is determined uniquely. The exact definition 
is \cite{kol1}
{\em
\par
The algorithmic complexity $K(x\mid y)$ of the  object $x$  by 
given object $y$ is the minimal length of the ``program'' $P$
that is written as a sequence of  the  zeros  and  unities
which allows us to construct $x$ having $y$:
\begin{equation}
K(x\mid y) = \min_{A( P,y)=x} l(P)
\label{cs-19}
\end{equation}
where $l(P)$ is length of the  program $P$; $A(P,y)$  is  the
algorithm  calculating  object $x$, using  the  program $P$,
when the object $y$ is given.
}
\par 
In this connection we can recall that 
't Hooft in Ref.\cite{hooft90} has proposed to investigate the Universe 
as a certain computer: 
``The finiteness of entropy of a black hole implies that the number of bits 
information that can be stored there is finite and determinated 
by the area of its horizon. This gave us the idea that Nature at the 
Planck scale is an information processing machine like a computer, 
or more precisely, a cellular automaton''.
\par
Now we can presuppose that the fluctuations of the metric signature 
occurs as the fluctuations between the algorithms (Einstein 
equations with the different metric signature)
\begin{equation}
\begin{array}{ccc}
\sigma = +1 & \longleftrightarrow & \sigma = -1
\\
& \Downarrow  & 
\\
G^+_{\bar 0\bar 0} 
& 
\longleftrightarrow
& 
G^-_{\bar 0\bar 0} 
\\
G^+_{\bar 1\bar 1} 
& 
\longleftrightarrow
& 
G^-_{\bar 1\bar 1}
\\
G^+_{\bar 2\bar 2} 
& 
\longleftrightarrow
& 
G^-_{\bar 2\bar 2}
\\
G^+_{\bar 3\bar 3} 
& 
\longleftrightarrow
& 
G^-_{\bar 3\bar 3}
\\
R^+_{\bar 5\bar 5} 
& 
\longleftrightarrow
& 
R^-_{\bar 5\bar 5}
\end{array}
\label{cs-20}
\end{equation}
The signs $\pm$ denote the belonging of the appropriate 
equation to the 
Euclidean or Lorentzian mode. The expression (\ref{cs-20}) 
designates that the appearance of the 
quantum magnitude $\sigma$ leads to a 
quantum fluctuation 
$R^+_{\bar A\bar B} \leftrightarrow R^-_{\bar A\bar B}$ 
or $G^+_{\bar A\bar B} \leftrightarrow G^-_{\bar A\bar B}$. 
The question is: how we can calculate a probability for 
each $R^\pm_{\bar A\bar B}$ ($G^\pm_{\bar A\bar B}$) 
equation? Our assumption for 
these calculations is that 
\textit{these probabilities are connected 
with the AC of each equation.} 

\subsection{Fluctuation 
$G^+_{\bar 2\bar 5} \longleftrightarrow G^-_{\bar 2\bar 5}$.} 

The $R_{\bar 2\bar 5}$ equation in the Euclidean mode is 
\begin{equation}
\ddot \omega + \dot \omega\left ( \frac{\dot a}{a} - 
\frac{\dot b}{2b} + 3\dot \psi \right )
-\frac{b}{a^2} \omega 
= 0
\label{ym1}
\end{equation}
and in the Lorentzian mode is 
\begin{equation}
\ddot \omega + \dot \omega\left ( \frac{\dot a}{a} - 
\frac{\dot b}{2b} + 3\dot \psi \right )
+ \frac{b}{a^2} \omega 
= 0
\label{ym2}
\end{equation}
Let we consider $\psi = 0$ case\footnote{below we 
will see that it will be in agreement with 
$R_{\bar 5\bar 5}$ equation}. It is easy to see that the first case 
can be deduced from the instanton condition 
\begin{equation}
E^2_1 = H^2_1 \qquad or \qquad 
\frac{\omega}{a} = \pm \frac{\dot\omega}{\sqrt b}
\label{ym3}
\end{equation}
The second equation (\ref{ym2}) have not such reduction 
from the instanton condition (\ref{ym3}) to this field equation. 
This is well known fact that the instanton 
can exist only in the Euclidean space. It allow us to say 
that the Euclidean equation (\ref{ym1}) is simpler 
from the algorithmical point of view than the Lorentzian 
equation (\ref{ym2}). In the first rough approximation 
we can suppose that the probability of the Euclidean 
mode is $p^+_{25} = 1$ and consequently for 
the Lorentzian mode $p^-_{25} = 0$. 
Strictly speaking the exact definition for 
each $p^\pm_{ab}$ probability should be \cite{dzh4} 
\begin{equation}
p^\pm_{ab} = \frac{e^{-K^\pm_{ab}}}
{e^{-K^+_{ab}} + e^{-K^-_{ab}}}
\label{ym4}
\end{equation}
here $K^\pm_{ab}$ is the AC for the $R^\pm_{ab} = 0$ 
equation. If $K^+_{25} \ll K^-_{25}$ then we have 
$p^+_{25} = 1$ and $p^-_{25} = 0$. 

\subsection{Fluctuation 
$R^+_{\bar 5\bar 5} \longleftrightarrow R^-_{\bar 5\bar 5}$.}

The $R_{\bar 5\bar 5}$ equation in the Euclidean mode is 
\begin{equation}
\ddot \psi + {\dot\psi}^2 + 
\frac{\dot a}{a}\dot\psi + \frac{\dot b}{b}\dot\psi + 
\frac{r_0^2}{2}e^{2\psi}\left (
H^2_{\bar 1} + E^2_{\bar 1}
\right ) =  0 ,
\label{ps1}
\end{equation}
and in the Lorentzian mode 
\begin{equation}
\ddot \psi + {\dot\psi}^2 + 
\frac{\dot a}{a}\dot\psi + \frac{\dot b}{b}\dot\psi + 
\frac{r_0^2}{2}e^{2\psi}\left (
-H^2_{\bar 1} + E^2_{\bar 1}
\right ) =  0 ,
\label{ps2}
\end{equation}
It is easy to see that the second equation (\ref{ps2}) (Lorentzian mode) 
is much more simpler as it has the trivial solution 
\begin{equation}
\psi = 0
\label{ps3}
\end{equation}
provided that we have the instanton condition 
\begin{equation}
H^2_{\bar 1} = E^2_{\bar 1}.  
\label{ps4}
\end{equation}
In the contrast with the previous 
subsection we see that in this case 
the Lorentzian mode is more preferable. As well in the 
contrast of the previous case we can suppose that in 
the first rough approximation the probability of the 
Euclidean mode of this equation is $p^+_{55} = 0$ and 
consequently for the Lorentzian mode $p^-_{55} = 1$ . 

\subsection{Fluctuation 
$G^+_{\bar 1\bar 1} \longleftrightarrow G^-_{\bar 1\bar 1}$ 
and 
$G^+_{\bar 2\bar 2} \longleftrightarrow G^-_{\bar 2\bar 2}$}

Taking into account (\ref{ps3}) we can write these equations 
in the following form 
\begin{eqnarray}
4\frac{\ddot a}{a} + 
\sigma\left (3\frac{b}{a^2} - \frac{4}{a} \right ) - 
\frac{{\dot a}^2}{a^2} + 
r_0^2e^{2\psi}\left (\sigma H_{\bar 1}^2 - E_{\bar 1}^2 
\right ) & = & 0 ,
\label{rm1}\\
2\frac{\ddot b}{b}  - 
\frac{\dot b^2}{b^2} + 2\frac{\ddot a}{a} + 
\frac{\dot a\dot b}{ab} - \frac{\dot a^2}{a^2} - 
\sigma\frac{b}{a^2} - 
r_0^2e^{2\psi}\left (\sigma H_{\bar 1}^2 - E_{\bar 1}^2 
\right ) & = & 0 ,
\label{rm2}
\end{eqnarray}
In the Euclidean mode ($\sigma = +1$) with the 
instanton condition (\ref{ps4}) it can be $b = a$ 
(an isotropical Universe) and we have \textit{only} 
one equation
\begin{equation}
4\frac{\ddot a}{a} - \frac{{\dot a}^2}{a^2} - \frac{1}{a} = 0 .
\label{rm3}
\end{equation}
In the Lorentzian mode ($\sigma = -1$) $b \ne a$ 
(this is a case of an nonisotropical Universe) and we have 
two equations 
\begin{eqnarray}
4\frac{\ddot a}{a} - 
\left (3\frac{b}{a^2} - \frac{4}{a} \right ) - 
\frac{{\dot a}^2}{a^2} - 
r_0^2e^{2\psi}\left (H_{\bar 1}^2 + E_{\bar 1}^2 
\right ) & = & 0 ,
\label{rm4}\\
2\frac{\ddot b}{b}  - 
\frac{\dot b^2}{b^2} + 2\frac{\ddot a}{a} + 
\frac{\dot a\dot b}{ab} - \frac{\dot a^2}{a^2} + 
\frac{b}{a^2} + 
r_0^2e^{2\psi}\left (H_{\bar 1}^2 + E_{\bar 1}^2 
\right ) & = & 0 ,
\label{rm5}
\end{eqnarray}
We see that in the Lorentzian mode we have the nonisotropical Universe 
as the consequence of the presence of the magnetic field 
$H_1 = \omega /a$. 
\par
Certainly the one equation in the Euclidean mode 
with the instanton condition (\ref{ps4}) is simpler 
than two equations in the Lorentzian mode. Our previous 
arguments according to the connection between the probability 
and the Kolmogorov's algorithmical complexity permit us 
to assume that in the first rough approximation the probability 
for (\ref{rm3}) in the Euclidean mode is $p^+_{11} = 1$ and 
consequently $p^-_{11}= 0$. 

\subsection{Fluctuation 
$G^+_{\bar 0\bar 0} \longleftrightarrow 
G^-_{\bar 0\bar 0}$}

Let we write the equation 
$G^\pm_{\bar 0\bar 0} = 0$ in the following form
\begin{equation}
2\frac{\dot b \dot \psi}{b} + 4 \frac{\dot a \dot \psi}{a} + 
2\frac{\dot a \dot b}{ab} + \frac{\dot a^2}{a^2} + 
\sigma \left (- \frac{4}{a} + \frac{b}{a^2} \right ) + 
r_0^2e^{2\psi}\left (\sigma H_{\bar 1}^2 - E_{\bar 1}^2 
\right ) = 0
\label{c1}
\end{equation}
By the conditions $\psi = 0$, instanton condition 
and $b = a$ 
this equation in the Euclidean mode is
\begin{equation}
\frac{\dot a^2}{a^2} - \frac{1}{a} = 0 
\label{c1a} 
\end{equation}
and in the Lorentzian mode
\begin{equation}
3\frac{\dot a^2}{a^2} + 3 \frac{1}{a} -
r_0^2e^{2\psi}\left (H_{\bar 1}^2 + E_{\bar 1}^2 
\right )= 0 .
\label{c2}
\end{equation}
Certainly the first Euclidean equation is simpler 
than the second Lorentzian one. In the first rough 
approximation this allow us to put $p^+_{00} = 1$ 
and $p^-_{00} = 0$. 

\subsection{Mixed system of the equations}

As the probability for each equations (\ref{cs-20}) are
only $p=0,1$ we can write the \textit{mixed} equation system for 
the Universe \textit{fluctuated between Euclidean and 
Lorentzian modes}
\begin{eqnarray}
\frac{\dot a^2}{a^2} - \frac{1}{a} & = & 0 , 
\label{mxo}\\
\dot \omega & = & \pm \frac{\omega}{\sqrt a} ,
\label{mx1}\\
4\frac{\ddot a}{a} - \frac{{\dot a}^2}{a^2} 
- \frac{1}{a} = 0 .
\label{mx2}
\end{eqnarray}
here $b = a$, $\psi = 0$ and the instanton condition (\ref{ps4}) 
is applied. It is easy to find the following solution
\begin{eqnarray}
a & = & \frac{t^2}{4} ,
\label{mx3}\\
\omega & = & t^2 .
\label{mx4}
\end{eqnarray}

\subsection{The mixed origin of the Universe}
\label{mixed}

\begin{figure}
\centerline{
\framebox{
\psfig{figure=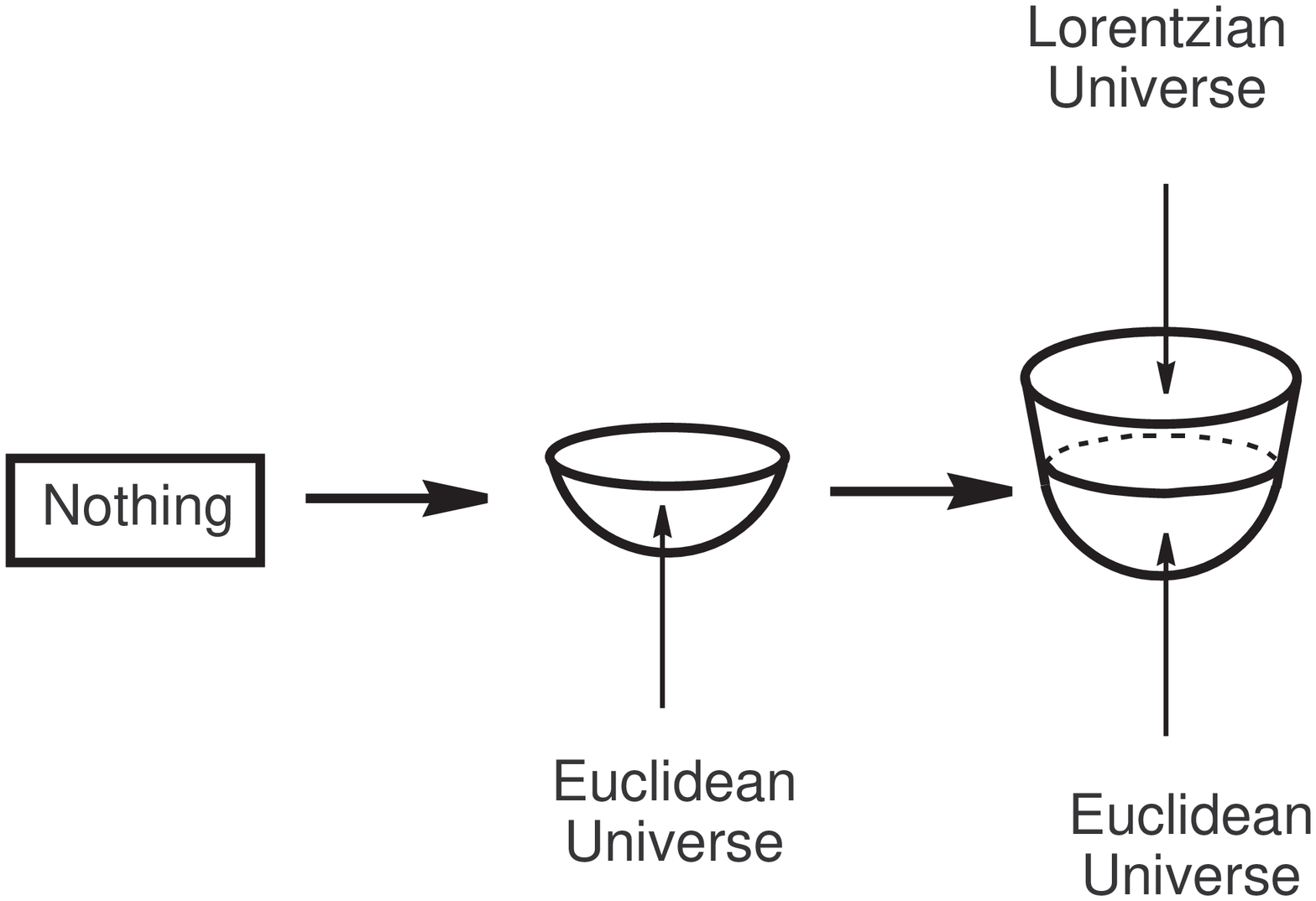,height=5cm,width=5cm}}}
\vspace{5mm}
\caption{An evolution of Lorentzian Universe from the ``Nothing''.}
\label{fig3}
\end{figure}

Hawking offers the following model of the quantum birth of 
Universe: at first appears a small piece of an Euclidean space 
($R^4$, $S^4$ or another smooth non-singular Euclidean space) 
then a Lorentzian Universe starts from a boundary of the 
initial Euclidean piece. In this scenario there is a hypersurface 
between Euclidean and Lorentzian spaces, see Fig.\ref{fig3}. 
\par
In this section we 
assume that at first there is a quantum Universe fluctuated 
between Euclidean and Lorentzian modes and later happens 
a quantum transition to the Lorentzian mode. 
\par
The 4D metric part of the MD metric with (\ref{mx3}) is 
\begin{equation}
ds^2_{(4)} = d\tau^2 + \tau^2d\Omega ^2_3
\label{eu5}
\end{equation}
where $\tau$ is the Euclidean time and 
$d\Omega ^2_3 = \frac{1}{4}[(d\xi + \cos \theta d\varphi )^2 + 
(d\theta ^2 + \sin ^2\theta d\varphi ^2)]$ is the metric 
of the 3D unit sphere $S^3$. The (\ref{eu5}) metric 
is the metric for the flat $R^4$ Euclidean space. 
\par
\begin{figure}
\centerline{
\framebox{
\psfig{figure=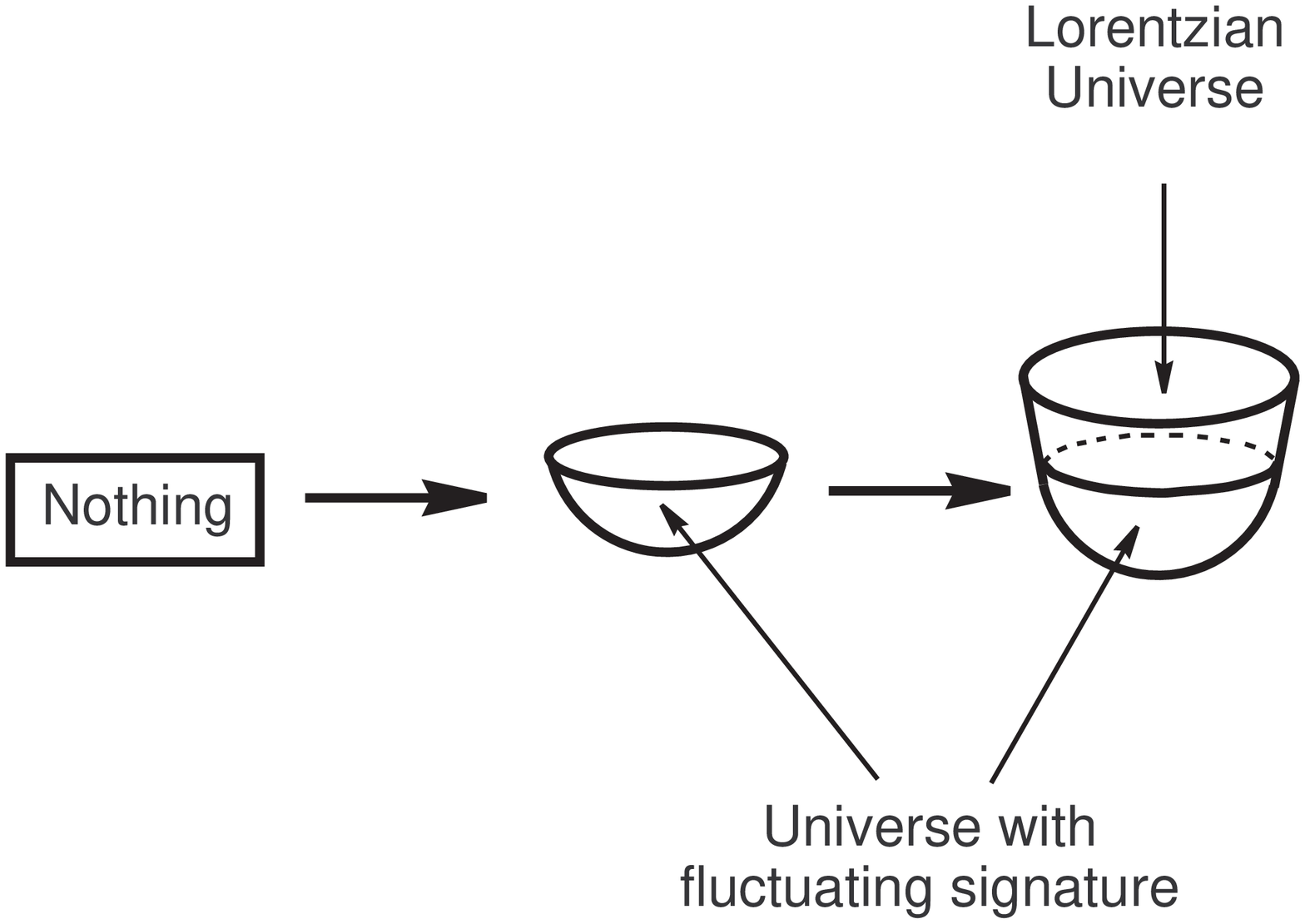,height=5cm,width=5cm}
}}
\vspace{5mm}
\caption{An evolution of Lorentzian Universe through the 
non-singular mixed Universe with the fluctuating metric signature.}
\label{fig6}
\end{figure}

Now we can assume that the fluctuating\footnote{between 
Euclidean and Lorentzian modes}, 
non-singular, multidimensional, empty 
Universe is born from ''Nothing`` as a piece 
($\tau \lesssim \tau_{Pl}$) 
of the $R^4 \times S^1$ Euclidean~$\leftrightarrow$~Lorentzian  
space. Then in some moment a quantum transition 
to one mode (Lorentzian) takes place and simultaneously 
(or later) the 
$G_{55}$ metric component become the non-dynamical variable. 
All this gives us the 4D Lorentzian Universe, 
see Fig.\ref{fig6}. 
\par 
\par 
Thus, the piece of the solution by $\tau \lesssim \tau_{Pl}$ 
can be interpreted as the quantum birth of the Universe 
with the fluctuating signature of metric. Roughly speaking, 
the 4D Universe\footnote{the base of the principal bundle} 
is in the Euclidean mode 
but the 5$^{th}$ extra dimension\footnote{the fibre of the 
principal bundle} is in the Lorentzian 
mode. As well we have the next interesting result: 
\textit{the fluctuation of the metric signature leads to 
frozing the part of MD metric connected with the 5$^{th}$ dimension}. 

\section{Conclusions}

In this paper we consider two examples of the 
nondifferentiable dynamic: the interchange of the 
sign between two components of the multidimensional 
metric and the quantum fluctuation between Euclidean and 
Lorentzian metrics. The existence of such kind of the 
nondifferentiable dynamic can lead to the interesting 
physical consequences: appearance of the spin-like structure 
on the throat of composite wormhole (this can be a possible 
connection with a superstring attached to two D-branes), 
frozing the physical degrees of freedom connected 
with the metric on the extra dimensions, quantum 
birth of the non-singular flat Universe with 
matter\footnote{gauge field as the non-diagonal components 
of the MD metric} and with the quantum fluctuation 
of the metric signature between Euclidean and 
Lorentzian modes. 
\par 
It is necessary to note that all these possibilities occur 
in a vacuum in the spirit of the Einstein idea that 
\textit{the right of the gravitational equations should be zero.} 
\par 
The Planck length paradigm says us that on this level 
the various unusual\footnote{from the viewpoint of the 
non-Planckian physic} 
phenomena can occur. In the application to our case this 
permits us to say that the discrete (nondifferentiable) 
quantities can fluctuate on this level\footnote{we can name 
the dynamic of 
these quantities as the nondifferentiable dynamic.}. 
It is possible that the physical phenomena connected with 
the nondifferentiable dynamic can play very important 
role in the very Early Universe or on the level of the 
spacetime foam. 
\par 
The basic idea presented by 't Hooft in Ref.\cite{hooft99} 
can be perceived so, that  
the fundamental states on the Planck level are the classical 
states and then the quantization should be stochastical: 
``$\ldots$ In our theory, quantum 
states are not the primary degrees of freedom. The primary 
degrees of freedom are deterministic states $\ldots$''. 
It is possible that the idea presented here have some 
connection with the 't Hooft stochastical quantization 
model. Actually, 
in the first case (section \ref{without}) we have the 
quantum fluctuation between two classical states 
(between two metric signatures 
$\eta _{ab} = (\pm ,-,-,-,\mp)$), in the second one 
we have the quantum fluctuating quantity $\sigma$ in the 
classical Einstein equations that leads to the fluctuation 
between Euclidean and Lorentzian modes with the 
stochastical definition of probability according to 
the (\ref{ym4}) equation.

\section{Acknowledgments}

This work is supported by a Georg Forster Research Fellowship
from the Alexander von Humboldt Foundation. I would like to
thank H.-J. Schmidt for the invitation to
Potsdam Universit\"at for research and D. Singleton 
for discussion.


\begin{thebibliography}{10}

\bibitem{ellis95}
M. Carfora and G. Ellis, Int. J. Mod. Phys. {\bf D4},  175  (1995).

\bibitem{ellis97}
C. Hellaby, A. Sumeruk, and G.~F.~R. Ellis, Int. J. Mod. Phys. {\bf D6},  211
  (1997).

\bibitem{dzh2}
V. Dzhunushaliev, Gen. Relat. Grav. {\bf 30},  583  (1998).

\bibitem{dzh7}
V. Dzhunushaliev, Mod. Phys. Lett. A {\bf 13},  2179  (1998).

\bibitem{wheel2}
C. Misner and J. Wheeler, Ann. Phys. {\bf 2},  525  (1957).

\bibitem{hooft99}
G. 't~Hooft, Class. Quant. Grav. {\bf 16},  3263  (1999).

\bibitem{Sal1}
A. Salam and J. Strathdee, Ann. Phys. {\bf 141},  316  (1982).

\bibitem{Per1}
R. Percacci, J. Math. Phys. {\bf 24},  807  (1983).

\bibitem{dzh4}
V. Dzhunushaliev, Class. Quant. Grav. {\bf 15},  603  (1998).

\bibitem{kol1}
A. Kolmogorov, {\em "Information Theory and Algorithm Theory: Logical
  Foundation of the Information Theory. Combinatorial Foundation of the
  Information Theory and the Probability Calculus"} (Nauka, Moscow, 1987).

\bibitem{hooft90}
G. 't~Hooft, Nucl. Phys. B {\bf 342},  471  (1990).

\end{thebibliography}

\end{document}